\documentclass[aps, pra, a4paper, showpacs, twocolumn, english, 10pt, nofootinbib]{revtex4-2}
\usepackage[T1]{fontenc}
\usepackage{babel}
\usepackage{bbm, amsthm, bm, textcomp, nicefrac, geometry, ragged2e}
\usepackage[dvipsnames]{xcolor}
\usepackage{float}
\usepackage[bbgreekl]{mathbbol}
\usepackage{graphicx, epstopdf, color, verbatim, enumitem, ulem}
\usepackage{pbox, array}
\usepackage{mathrsfs}
\usepackage{physics}
\usepackage{amssymb}
\usepackage{mathtools}
\usepackage{stackrel}
\usepackage[thinlines]{easytable}
\usepackage{amsmath}
\usepackage{makecell}
\usepackage{aligned-overset}

\makeatletter
\usepackage[caption=false]{subfig}
\addto\captionsenglish{}
\usepackage{hyperref}
\usepackage{tikz}
\usetikzlibrary{shapes,arrows, positioning, calc}  

\hypersetup{
    colorlinks = true,
    linkcolor = BrickRed,
    citecolor = BrickRed,
    filecolor = Black,     
    urlcolor = Black,
}
\DeclareMathSymbol{\shortminus}{\mathbin}{AMSa}{"39}

\renewcommand\bra[1]{{\langle{#1}|}}
\renewcommand\ket[1]{{|{#1}\rangle}}
\begin{document}

\title{On the power of moving quantum sensors: fully flexible and noise-resilient sensing}
% / P:Quantum metrology moving beyond static limits}
\author{Paul Aigner}
\email[Corresponding author: ]{paul.aigner@uibk.ac.at}
\affiliation{Institut f\"ur Theoretische Physik, Universit\"at Innsbruck, Technikerstra{\ss}e 21a, 6020 Innsbruck, Austria}
\author{Wolfgang D\"ur}
\affiliation{Institut f\"ur Theoretische Physik, Universit\"at Innsbruck, Technikerstra{\ss}e 21a, 6020 Innsbruck, Austria}
\date{\today}

%% Abstract %%
\begin{abstract}
We show that a single moving quantum sensor provides complete access to spatially correlated scalar fields. We demonstrate that with either trajectory or internal state control, one can selectively measure any linear functional, e.g. a gradient or a spatial Fourier series coefficient, while successfully eliminating {\it all} noise signals with orthogonal spatial correlation. This even exceeds the capabilities of a sensor network consisting of multiple entangled, yet spatially fixed, quantum sensors, where the number of suppressed noise signals is limited by the number of sensor positions. We show that one can achieve an improved scaling of the quantum Fisher information for moving sensors beyond the static fundamental limit of $T^2$.
\end{abstract}

%%% Title %%
\maketitle
\predisplaypenalty=10000
\postdisplaypenalty=10000
%% Body %%

\textit{Introduction.—}
Quantum metrology exploits quantum mechanical effects to estimate parameters with precision beyond classical limits~\cite{Giovannetti2011,Pezzè2018,Degen2017}. In settings where the imprinting Hamiltonian is time independent, the optimal precision is determined by the quantum Fisher information (QFI) which has the familiar $T^2$ time scaling limit per probe, with entanglement offering Heisenberg scaling $N^2$ in the sensor number $N$ when noise is absent~\cite{Giovannetti2011,Pezzè2018}. Recently, progress has been made on using distributed probes across networks to sense spatially varying fields~\cite{Komar2014,Eldredge2018,PhysRevA.103.L030601}, and on mitigating decoherence with techniques ranging from decoherence-free subspaces (DFS)~\cite{PhysRevResearch.2.023052,Hamann_2024} to quantum error correction (QEC) and fast control~\cite{Kessler2014,PhysRevLett.112.080801,Demkowicz2017,Sekatski2017quantummetrology}.

However, most current approaches assume that sensors remain in fixed positions and, hence, their output is determined only by the field values at those stationary points. This perspective implicitly constrains both which functionals of a field can be accessed and how many independent noise contributions can be cancelled~\cite{PhysRevResearch.2.023052,PhysRevA.103.L030601}. 

Here, in contrast, we consider mobile quantum sensors whose position and internal dynamics can be shaped by designing a trajectory and quantum control, respectively. We demonstrate that the imprinted phase becomes an engineered linear functional of the field along the path, which enables selective readout of arbitrary linear functionals (e.g., gradients or a spatial Fourier component) and the cancelation of any number of fluctuating signals whose spatial correlations are linearly independent of the signal. Those are capabilities that strictly exceed those of static networks with the same number of physical qubits.

Mobility also fundamentally changes the metrological time scaling. Motion makes the imprinting Hamiltonian time dependent, where with adaptive control one can surpass the static $T^2$ scaling limit of the QFI \cite{Pang2017}. We show this mechanism with a minimal moving sensor example and demonstrate that a moving sensor can have an increased QFI growth rate. 

In addition, we compare mobile sensing against the best static network strategies for dealing with fluctuating noise signals. Even in regimes where networks can cancel noise, a single $N$-qubit sensor that sequentially visits the $N$ positions of a fixed position network can achieve a strictly larger QFI than the sensor network. Finally, we extend our constructions to fields with explicit time dependence, showing how adaptable velocity schedules and quantum control decouple spatial weighting from temporal reshaping, enabling noise cancelation.

Moving quantum systems where the quantum information carriers can be physically transported or rearranged over short distances in a controlled way are natively available on several  quantum computing platforms. In certain trapped ion processors, one can shuttle ions between zones in segmented traps \cite{Pino2021,PhysRevX.13.041052,Akhtar2023}. Furthermore, neutral atom computers similarly allow atoms to be repositioned using optical tweezers, making large-scale rearrangements possible \cite{Chiu2025,Bluvstein2024,PhysRevX.12.011040,wurtz2023aquilaqueras256qubitneutralatom,Barredo2018}. Such systems can also be used as quantum sensors, and hence our approach not only promises theoretical advantages but could be readily implemented in existing platforms.~Other conceivable examples of moving sensors include quantum sensors mounted on satellites, vehicles, or drones. While for the former trajectories are largely fixed, the latter offer also trajectory control.  

Taken together, these results establish motion as a powerful and largely under-explored resource for quantum metrology. By designing trajectories and quantum control, one gains full access to linear functionals of spatially (and temporally) correlated signals, achieves robust noise cancelation beyond static limits, and enables QFI time scaling inaccessible to stationary sensors.

\textit{Background.—}
We briefly summarize the main theoretical concepts and figures of merit for single-parameter quantum metrology in the Fisher regime, i.e. the sensing regime of many $m\gg 1$ measurements \cite{Helstrom1969}. Although most of our results are valid beyond the Fisher regime, we adopt it here for the clarity of presentation and use the QFI as the primary figure of merit. In single-parameter quantum metrology, the objective is to estimate an unknown parameter $\omega$ encoded into a quantum state through an imprinting unitary operation
$
U(\omega) = \exp(-it G \omega),
$
where $G$ is the generator of the parameter-dependent evolution. The precision of any unbiased estimator $\hat{\omega}$ is fundamentally limited by the quantum Cramér--Rao bound, which relates the mean squared error to the QFI as 
$
\mathrm{Var}(\hat{\omega}) \ge (m {I}_\omega^{(Q)}[\rho(\omega)])^{-1}
$
 \cite{Helstrom1969}.
The QFI quantifies the sensitivity of a quantum state $\rho(\omega)$ to changes in $\omega$, and is defined for pure states $|\psi_\omega\rangle = U(\omega)|\psi_0\rangle$ by \cite{PhysRevLett.72.3439}
\begin{equation}
I^{(Q)}_\omega = 4t^2\left( \langle \psi_0 | G^2 | \psi_0 \rangle - \langle \psi_0 | G | \psi_0 \rangle^2 \right) = 4t^2 (\Delta G)^2.
\end{equation}
For $N$ probe states and interrogation time $T$ one finds that the QFI scaling is limited by $T^2N^2$ \cite{PhysRevLett.72.3439,PhysRevLett.96.010401}.

\begin{figure}[t]
    \centering
    \begin{minipage}{1\columnwidth}
    \centering
    \includegraphics[width=\linewidth]{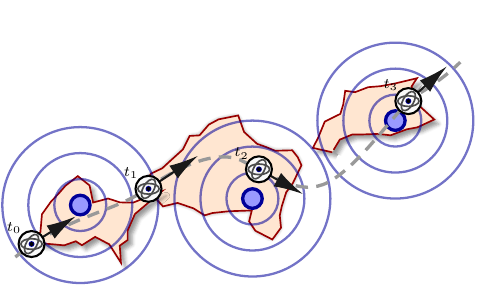}
    \vspace{0.3em}
    {\small (a)}
\end{minipage}

\begin{minipage}{1\columnwidth}
    \centering
    \includegraphics[width=\linewidth]{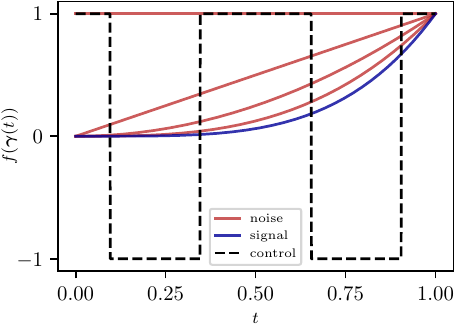}
    %\vspace{0.3em}
    {\small (b)}
\end{minipage}

    \caption{(a) Sketch of a mobile quantum sensor, moving through signal fields, illustrated by blue concentric circles, and noise fields, illustrated by jagged orange blobs. The quantum sensor follows a path, illustrated by the gray dashed line, where it has at different times $t_i$ different local velocities, illustrated with black arrows. (b) Plot of the signal generating function $f(\boldsymbol{\gamma}(t))=t^4$, noise generating functions $g_i(\boldsymbol{\gamma}(t)) \in \{t^0,t^1,t^2,t^3\}$, and corresponding optimal control function $\mathrm{sgn}[f(\boldsymbol{\gamma})-p^*](t)$, with $p^*$ being the best $L^1[0,1]$ approximation of $f(\boldsymbol{\gamma})$ out of $\mathrm{span}_{\mathbbm{R}}\{t^0,t^1,t^2,t^3\}$. The control function $\mathrm{sgn}[f(\boldsymbol{\gamma})-p^*]$ cancels the contributions of the noise signals $g_i({\boldsymbol{\gamma}})$, while maximizing the residual sensitivity to the signal function $f(\boldsymbol{\gamma})$.}
    \label{fig:1}
\end{figure}

\textit{Setting and general methodology.—} We consider the estimation of a signal described by the Hamiltonian of the form 
$
    H_{\text{signal}}= f(\boldsymbol{x}_{\text{sensor}}(t)) G, 
$
where $G$ is a hermitian generator with symmetric spectrum, $\boldsymbol{x}_{\text{sensor}}(t)$ is the position of the sensor at time $t$, and we want to estimate some property of the position dependent scalar field $f: \mathbbm{R}^d \rightarrow \mathbbm{R}$. Assuming our sensor position follows a path $\boldsymbol{\gamma}:[0,T] \rightarrow \mathbbm{R}^d$, we obtain the imprinting unitary
\begin{equation}
U_\Phi=\exp(-i\Phi G), \ \Phi=\int_0^T f(\boldsymbol{\gamma}(t))dt.
\end{equation}
We show that by appropriately controlling the sensor trajectory and phase imprinting via sign-control, any linear functional of the field $f(\boldsymbol{\gamma})$ can be realized up to scaling. Our methodology is state independent and therefore applicable in both Fisher and Bayesian estimation settings. Fixing the geometric path $\gamma([0,T])$, we first vary only the velocity schedule by introducing a smooth, strictly increasing reparametrization $h:[0,T]\rightarrow[0,T]$, yielding
\begin{equation}
    \Phi_h=\!\int_0^T f(\boldsymbol{\gamma}(h(t)))dt=\!\int_0^T f(\boldsymbol{\gamma}(\tau))w(\tau)d\tau,
\end{equation}
where $w(\tau)\equiv 1/h'(\tau) \ge 0$ satisfies $\int_0^T w(\tau)d\tau=T$. By the Riesz representation theorem~\cite{ASENS_1911_3_28__33_0}, as already indicated in \cite{PhysRevA.103.L030601}, appropriate choices of a velocity schedule allow realization of any non-negative linear functional of $f(\boldsymbol{\gamma})$. Introducing additional sign control $s(t)\in\{0,1\}$ gives
\begin{equation}
    \Phi_{w,s}=\int_0^T f(\boldsymbol{\gamma}(t))w(t)(-1)^{s(t)}dt,
\end{equation}
thus enabling the imprinting of arbitrary linear functionals.
Similarly, with a fixed velocity schedule but bounded control function $c(t)\!\in\![-1,1]$, the accumulated phase
\begin{equation}
    \Phi_c=\int_0^T f(\boldsymbol{\gamma}(t))c(t)dt
\end{equation}
again represents an arbitrary linear functional of $f(\boldsymbol{\gamma})$. If we consider analytical functions $f$, then varying trajectories on any open set and allowing for sign control suffice to encode an arbitrary non-negative linear functional of $f$, as analytical functions are uniquely defined via their values on any open set. Notice that this selective imprinting is different from inferring the value of a known functional from some function values, in particular when signals with different functional dependencies are considered.  

\textit{Enhancing noiseless estimation through time-dependent 
control.—} 
We show that moving quantum sensors can surpass the fundamental $T^2$ QFI scaling limit of time-independent metrology by converting a time-independent sensing task into a time-dependent one. This allows us to exploit adaptive control strategies for time-dependent imprinting Hamiltonians, which are known to beat the standard scaling~\cite{Pang2017}.
For a probe state $\ket{\Psi_0}$ evolving under a parameter-dependent channel $\Lambda_g$, the QFI
\begin{equation}
    I_g^{(Q)} = 4\big(\braket{\partial_g \Psi_g}{\partial_g \Psi_g} 
    - |\braket{\Psi_g}{\partial_g \Psi_g}|^2\big)
\end{equation}
quantifies the maximum information about $g$ encoded in 
$\ket{\Psi_g} = \Lambda_g\ket{\Psi_0}$.  
For a time-dependent imprinting Hamiltonian $H_g(t)$, one finds the general bound
\begin{equation} \label{eq:QFI_bound_main}
    I_g^{(Q)} \le \!\left[\! \int_0^T (\mu_{\max}(t)-\mu_{\min}(t))\,dt \!\right]^2,
\end{equation}
where $\mu_{\max,\min}(t)$ are the extremal eigenvalues of 
$\partial_g H_g(t)$.  
This bound can be saturated using an adaptive control Hamiltonian $H_c(t)$ that dynamically aligns the probe with the instantaneous eigenbasis of $H_g(t)$, achieving the optimal QFI scaling \cite{Pang2017}.
Hence, given a natively time-independent sensing problem, via introducing a moving sensor, we can make it a time-dependent sensing problem allowing for an enhanced QFI scaling.

As an illustrative example, consider estimating a spatial frequency $k$ in 
\begin{equation}
H_k(t)=-B[\cos(k \gamma(t))\sigma_x+\sin(k\gamma(t))\sigma_y],
\end{equation}
with a one-dimensional sensor path $\gamma(t)=vt$.  
Here, $\partial_k H_k(t)$ has eigenvalues $\pm vtB$, yielding 
$I_k^{(Q)} = B^2v^2T^4$.  
Thus, the obtainable precision scales quartically in time, in contrast to the $T^2$ scaling limit of static sensors.  
For general one-dimensional paths $\gamma(t)=\int_0^t v(\tau)\,d\tau$, the scaling can exceed $T^4$ and the QFI of the optimal velocity schedule is given by
$
    I_k^{(Q)}=4B^2T^2L^2,
$
where $L$ is the distance covered by the sensor.
Full derivations and extensions to arbitrary signal Hamiltonians are given in Appendix~\ref{app:timedep}. 

\textit{Noise cancelation.—}
Considering the more realistic scenario of the estimation of a signal in the presence of noise signals, we demonstrate that we can cancel \textit{all} noise signals which are linearly independent of the signal. Assume the imprinting Hamiltonian is given as
$
    H=\omega f(\boldsymbol{\gamma}(t))G+\sum_{j=1}^n \beta_j g_j(\boldsymbol{\gamma}(t))G,
$
where $\omega$ is the parameter of interest, $\beta_j$ are fluctuating amplitudes, $f$ and $\{g_j\}_{j=1}^n$ are linearly independent functions, and $\boldsymbol{\gamma}$ is the sensor path, see Fig.~\ref{fig:1}(a) for a sketch of the setting. As noise deteriorates the sensing performence we want to cancel their contributions
\begin{equation}
    \int_0^T g_j(\boldsymbol{\gamma}(t))c(t)dt=0, \ \forall j, 
\end{equation}
while maximizing the residual sensitivity $\Phi_c$.
% \begin{equation}
%     \Phi_{c}=\int_0^T f(\boldsymbol{\gamma}(t)) c(t)dt.
% \end{equation}
We seek a control function that nullifies the linear functionals of $  \{g_j(\boldsymbol{\gamma})\} $ while keeping that of $f(\boldsymbol{\gamma})$ non-zero.  
The functions $g_j(\boldsymbol{\gamma})$ span a finite-dimensional subspace
$
    V_n = \mathrm{span}_{\mathbbm{R}}\{ g_1(\boldsymbol{\gamma}), \dots, g_n(\boldsymbol{\gamma}) \} 
   ,
$
equipped with the inner product
\begin{equation}
    \langle \phi, \xi \rangle = \int_0^T \phi (t) \xi(t)\, dt.
\end{equation}
The orthogonal complement
$
   V_n^\perp = \{ \phi \mid \langle \phi, \xi \rangle = 0,\ \forall \xi \in V_n \}
$
contains all control functions that cancel all noise-induced phase contributions.  
Since
$
    (V_n^\perp)^\perp \subset V_n,
$
any signal component outside the linear span of $\{g_j(\boldsymbol{\gamma})\}$ can be isolated by a control in $V_N^\perp$, which cancels noise without removing the signal. Note that if $f$ and $\{g_j\}$ are linearly independent then there always exists a path $\boldsymbol{\gamma}$ on which they are also linearly independent, see Appendix~\ref{app:path_noise}. Hereafter, we assume that a path $\boldsymbol{\gamma}$ is given such that 
$f(\boldsymbol{\gamma})$ and $\{g_j(\boldsymbol{\gamma})\}$ are linearly independent.

The Hobby--Rice theorem~\cite{hobbyrice} asserts that for any $n$ real-valued integrable functions $ g_1(\boldsymbol{\gamma}),\dots,g_n(\boldsymbol{\gamma})$ on $[0,T]$, there exist partition points
$
0 = t_0 < t_1 < \cdots < t_n = T
$
and signs $\delta_j \in \{\pm1\}$ such that, for all $i=1,\dots,n$,
\begin{equation}
\sum_{j=1}^{n} \delta_j \int_{t_{j-1}}^{t_j} g_i(\boldsymbol{\gamma}(t))\, dt = 0.
\end{equation}
Hence, with at most $n+1$ sign control operations, one can cancel up to $n$ linearly independent noise terms.  

The above construction only asserts the existence of such a schedule, however, does not directly address the exact implementation schedule and is indifferent to the residual sensitivity. To address these points, we construct the optimal control schedule with respect to the residual sensitivity. To this aim, we define the linear combination 
$
    L(\boldsymbol{\alpha}) = \sum_{i=1}^n \boldsymbol{\alpha}_i g_i(\boldsymbol{\gamma}),
$
 and we call the function $L(\boldsymbol{\alpha}^*)$ the best $L^1$-approximation to $f(\boldsymbol{\gamma})$ out of $V_n$ if
$
    \|L(\boldsymbol{\alpha}^*) - f(\boldsymbol{\gamma})\|_1 \le 
    \|L(\boldsymbol{\alpha}) - f(\boldsymbol{\gamma})\|_1,
$
for all $\boldsymbol{\alpha}$.
We define the coincidence set 
$
    Z(\boldsymbol{\alpha}) = \{t \mid f(\boldsymbol{\gamma}(t)) = L(\boldsymbol{\alpha},t)\}.
$
Then a necessary and sufficient condition for $ L(\boldsymbol{\alpha}^*) $ to be the best $L^1$-approximation is~\cite{hobbyrice,7cd95434-5c8f-30be-9320-e9324d15721e,doi:10.1073/pnas.45.10.1523}
\begin{equation}
\begin{aligned}
    \left| \int_0^T L(\boldsymbol{\alpha},t)\,
    \mathrm{sgn}[f(\boldsymbol{\gamma}) - L(\boldsymbol{\alpha}^*)](t)\, dt \right|
    \le \int_{Z(\boldsymbol{\alpha}^*)} |L(\boldsymbol{\alpha},t)|\, dt
\end{aligned}
\end{equation}
for all $ \boldsymbol{\alpha} $.  
If the measure of $Z(\boldsymbol{\alpha}^*)$ is zero, this simplifies to
\begin{equation}
    \int_0^T L(\boldsymbol{\alpha},t)\,
    \mathrm{sgn}[f(\boldsymbol{\gamma}) - L(\boldsymbol{\alpha}^*)](t)\, dt = 0,
    \quad \forall\, \boldsymbol{\alpha}.
\end{equation}
Equivalently, the optimal control cancels all noise contribution whilst yielding the maximum residual sensitivity
\begin{equation}
\begin{aligned}
   s_{V_n,f}&= \int_0^T \left|\,[f(\boldsymbol{\gamma})- L(\boldsymbol{\alpha}^*)](t)\, \right |\ dt .
\end{aligned}
\end{equation}
If $f(\boldsymbol{\gamma})$ and $ \{g_i(\boldsymbol{\gamma})\} $ are analytic, then $ Z(\boldsymbol{\alpha}^*)$ has measure zero, since the zeros of non zero analytic functions form a set of zero measure~\cite{kuchment2016overviewperiodicellipticoperators,dang2015complexpowersanalyticfunctions,mityagin2015zerosetrealanalytic}. See Fig.~\ref{fig:1}(b) for an example of the methodology. In practice, the best $L^1$-approximation can be obtained numerically by evaluating the integrals on quadratures and solving the resulting linear program~\cite{abdelmalek1975efficient,barrodale1967approximation}. 

For certain families of basis functions, motivated by standard series expansions such as Taylor or Fourier series, it is possible to derive explicit constructions of both the control function and the resulting residual sensitivity. In particular, by designing the control to follow a function from an orthogonal basis on a given interval, for example, a cosine profile or a Legendre polynomial, one can cancel {\it all} basis functions orthogonal to the chosen signal. As a result, the system becomes insensitive to noise with any spatial profile that differs from the signal. An explicit construction of this procedure is given in Appendix~\ref{app:cancelling-noise-families}. 

Hence, we can cancel an arbitrary amount of noise signals, with a single sensor, in contrast to stationary sensor networks employing DFS, where the total number of noise signals that can be canceled is limited by the number of sensor positions~\cite{PhysRevResearch.2.023052}. Note that in this setting, DFS are as powerful as using quantum error correction with full control \cite{Hamann_2024}. 

\textit{Comparison to sensor networks.—} Even in the regime where sensor networks can cancel noise contributions, i.e. the number of noise signals is smaller than the number of sensor positions, we demonstrate via the example of a fast-moving sensor that one can outperform any static sensor network. We consider an idealized limit in which the sensor can be repositioned arbitrarily quickly. The imprinting Hamiltonian is
\begin{equation}
    H=\omega \sum_{i=1}^N f(\boldsymbol{\gamma}^{(i)}(t))\sigma_z^{(i)}
    +\sum_{j=1}^n \beta_j \sum_{i=1}^N g_j(\boldsymbol{\gamma}^i(t))\sigma_z^{(i)},
\end{equation}
where $\sigma_z^{(i)}$ are local Pauli-$Z$ operators. Note that the following construction and conclusion do not depend on the specific choice of commuting generators.

A static network can cancel the noise through a DFS only if $N\ge n+1$, provided the noise contributions at the sensor positions are linearly independent of the signal~\cite{PhysRevResearch.2.023052}. We compare a static sensor network of $N\ge n+1$ distributed qubits with a single sensor comprising $N$ qubits that can move instantaneously between the static sensor positions, where it spends a fraction of the total sensing time.

In the presence of noise, DFS strategies suppress decoherence by engineering destructive interference between noise contributions from different sensing locations. This is achieved by initially preparing a GHZ-type state $\frac{1}{\sqrt{2}}(\ket{0}^{\otimes N} + \ket{1}^{\otimes N})$ and tailoring effective local signal-coupling parameters such that all noise terms cancel, while retaining sensitivity to the signal of interest. The effective local signal-coupling parameters are implemented through local bit-flip gates applied at suitable intermediate times, where each bit flip reverses the sign of the local interaction, enabling controlled modulation of the effective coupling. Concretely, one introduces a set of coefficients $\boldsymbol{s}_i$ that modulate the local couplings, chosen to satisfy
$
    \sum_{i=1}^N g_j(\boldsymbol{x}^{(i)}) \boldsymbol{s}_i = 0,
$
for all noise sources $j$, while maximizing the residual signal. As shown in Ref.~\cite{PhysRevResearch.2.023052}, this construction defines a DFS in which decoherence is eliminated at the cost of a reduced effective signal strength. The resulting performance of an optimal static sensor network employing a DFS is quantified by the QFI,
\begin{equation}
    I_\omega^{(Q)} = 4\,\langle \boldsymbol{s}, \boldsymbol{s}^\star \rangle^2 T^2,
\end{equation}
where $\boldsymbol{s}=(f(\boldsymbol{x}^{(1)}),\dots,f(\boldsymbol{x}^{(N)}))$, and $\boldsymbol{s}^\star$ is the optimal choice of coefficients,
\begin{equation}
\boldsymbol{s}^\star
= \arg\!\max_{\boldsymbol{s}\in[-1,1]^N}
\left\{
\sum_{i=1}^N f(\boldsymbol{x}^{(i)}) \boldsymbol{s}_i
\;\bigg|\;
\sum_{i=1}^N g_j(\boldsymbol{x}^{(i)}) \boldsymbol{s}_i = 0,\ \forall j
\right\}.
\end{equation}

We now contrast this with the moving-sensor strategy. As in the DFS strategy, we initialize the probe state in a GHZ-type state
$
\frac{1}{\sqrt{2}}\!\left(\ket{0}^{\otimes N}+\ket{1}^{\otimes N}\right).
$
After an interrogation time $T$ a single sensor moving among the network sensor positions acquires a relative phase
\begin{equation}
    \phi
    = 2TN\!\sum_{j=1}^N r_j \left(f(\boldsymbol{x}^{(j)}) + \sum_{i=1}^n \beta_i g_i(\boldsymbol{x}^{(j)}) \right),
\end{equation}
with $r_j \ge 0$ denoting the fraction of time spent at position $\boldsymbol{x}^{(j)}$, satisfying $\sum_j r_j = 1$. The time weights $r_j$ in the moving-sensor protocol play a role analogous to the coefficients $\boldsymbol{s}_j$ in DFS-based networks. By appropriately choosing $r_j$ and interaction signs for the corresponding time periods, noise contributions can be averaged away. Unlike the DFS approach, however, this cancellation does not require reducing the effective signal amplitude. Specifically, setting
$
    r_j = \frac{|\boldsymbol{s}_j^\star|}{\|\boldsymbol{s}^\star\|_1},
$
and implementing the signs $\mathrm{sgn}(\boldsymbol{s}_j^\star)$ via local sign control, the moving sensor achieves a QFI
\begin{equation}
    I_\omega^{(Q)}
    = 4\!\left(\frac{N}{\|\boldsymbol{s}^\star\|_1}\right)^2
    \langle \boldsymbol{s}, \boldsymbol{s}^\star \rangle^2 T^2.
\end{equation}
This corresponds to an enhancement factor of
$
    \left(N/\|\boldsymbol{s}^\star\|_1\right)^2.
$

Thus, a fast-moving sensor can achieve a QFI that is equal to or larger than that of any static network, while also avoiding the need to distribute entanglement across distant sensor stations, since the initial GHZ state can be prepared locally on a single device.

\textit{Spatially and temporally correlated signals.—}
We now generalize the previous time-independent signal model to signals with explicit space and time dependence.  
Let the signal Hamiltonian be
$
    H_{\mathrm{signal}}(t) = f(\boldsymbol{x}_{\mathrm{sensor}}(t), t)\, G,
$
where
$f(\boldsymbol{x},t)$ encodes both spatial and temporal variations of the field.  
For a given sensor path $\boldsymbol{\gamma}$, the accumulated phase is given by
\begin{equation}
    \Phi = \int_0^T f(\boldsymbol{\gamma}(t), t)\, dt.
\end{equation}
To explore how motion affects phase accumulation, consider a smooth, bijective re-parametrization 
$h:[0,T]\!\to\![0,T]$ that modifies the velocity profile along the same geometric path $\boldsymbol{\gamma}([0,T])$.  
The re-parametrized phase becomes
\begin{equation}
    \Phi_h = \int_0^T 
    f(\boldsymbol{\gamma}(\tau), h^{-1}(\tau))\, 
    w_h(\tau)\, d\tau,
\end{equation}
where $w_h=1/h'$ acts as an effective time-dependent weight induced by the re-parametrization.  
Thus, modifying the sensor’s speed profile introduces both a weighting function $w_h$ and an explicit reshaping of the time dependence $t \mapsto h^{-1}(t)$.
By introducing an additional quantum control function 
$\tilde{c}(t) = c(t)/w_h(t)$,
we can independently modulate the weighting and temporal structure of the accumulated phase,
\begin{equation}
    \Phi_{c,h} = \int_0^T 
    f(\boldsymbol{\gamma}(t), h^{-1}(t))\, c(t)\, dt.
\end{equation}
This decouples the spatial motion control from the explicit time-dependent control, 
while constraining the admissible amplitudes of the time-dependent control via 
$\|\tilde{c}\|_\infty \le 1$.  

Considering noisy signals, the imprinting Hamiltonian is given as
$
    H=\omega f(\boldsymbol{\gamma}(t),t)G+\sum_{j=1}^n \beta_j g_j(\boldsymbol{\gamma}(t),t)G,
$
where $f$ and $\{g_j\}$ are linearly independent functions. If $f(\boldsymbol{\gamma}(\cdot),\cdot)$ and $\{g_j(\boldsymbol{\gamma}(\cdot),\cdot)\}$ are linearly independent and have a coincidence set of measure zero, then analogous to the time-independent case we can find the optimal sign-control sequence yielding 
\begin{equation}
    \int_0^T g_j(\boldsymbol{\gamma}(t),t)\mathrm{sgn}[f(\boldsymbol{\gamma}(\cdot),\cdot)-L(\boldsymbol{\alpha^*})](t) dt = 0, 
\end{equation}
for all $j$, where $L(\boldsymbol{\alpha^*})$ is the best $L^1([0,T])$ approximation of $f(\boldsymbol{\gamma}(\cdot),\cdot)$ out of the linear span of $\{g_j(\boldsymbol{\gamma}(\cdot),\cdot)\}$, as well as
obtaining the optimal residual sensitivity
\begin{equation}
    s_{f,V_n}=\int_0^T f(\boldsymbol{\gamma}(t),t)\mathrm{sgn}[f(\boldsymbol{\gamma}(\cdot),\cdot)-L(\boldsymbol{\alpha^*})](t) dt.
\end{equation}
As in the time independent case, if $f$ and $\{g_j\}$ are linearly independent, then there always exists a path $\boldsymbol{\gamma}$ on which they are also linearly independent, see Appendix~\ref{app:noisecanctime}.

\textit{Conclusion.—} We have demonstrated that with a single moving quantum sensor one can cancel the contribution of an arbitrary amount of fluctuating scalar fields, which are linearly independent of the signal of interest. For a given sensor path, we constructed the optimal sign-control sequence, which maximizes the residual sensitivity of the signal, while canceling all noise contributions. Through the example of a fast-moving sensor, we have shown that any quantum sensor network with fixed sensor positions, which utilizes DFS to cancel noise contributions, can be outperformed. Furthermore, we have seen that we can turn natively time-independent sensing problems to effective time-dependent ones, which allows us to surpass the fundamental time scaling limit of static sensors. Finally, we have shown that we can extend our results to space- and time dependent scalar fields. 

We have shown the potential of a single moving quantum sensor to estimate a single parameter of interest. Possible extensions of our approach include generalizations to multi-parameter sensing problems, non-commuting generators and the investigation of a network of multiple moving sensors. For stationary sensor, such sensor networks allow for metrological capabilities not present in a single sensor, and it will be interesting to see if networks of moving sensors offer additional advantages and capabilities. Overall, our findings show that the investigation of mobile quantum sensors has the potential to enhance quantum metrology in diverse settings. 

\textit{Acknowledgments.—}
The authors acknowledge support from the Austrian Science Fund (FWF). This research was funded in whole or in part by the Austrian Science Fund (FWF) 10.55776/P36009, 10.55776/P36010 and 10.55776/COE1. For open access purposes, the author has applied a CC BY public copyright license to any author accepted manuscript version arising from this submission. Finanziert von der Europäischen Union.

\bibliographystyle{apsrev4-1}

\bibliography{ref.bib}
%% Appendix
\renewcommand\appendixname{Appendix}
\appendix

\section{Cancelling of noise through path design}\label{app:path_noise}

Given signal function $f: \mathbbm{R}^d \rightarrow \mathbbm{R}$, and noisy functions $\{g_i:  \mathbbm{R}^d \rightarrow \mathbbm{R}\}_{i=1}^n$. Here, we show that assuming $f$ and $\{g_i\}$ are linearly-independent, there exists a path $\boldsymbol{\gamma}:[0,T] \rightarrow \mathbbm{R}^n$ such that $f(\boldsymbol{\gamma})$ and $\{g_i(\boldsymbol{\gamma})\}_{i=1}^n$ are linearly independent. 
\textbf{Proof:} Given $m \equiv n+1$ linearly independent functions $\{f_i:\mathbbm{R}^d\rightarrow \mathbbm{R}\}_{i=1}^m$
Define $F:\mathbbm{R}^d\to\mathbbm{R}^m$ by
\begin{equation}
\boldsymbol{F}(\boldsymbol{x}) := \big(f_1(\boldsymbol{x}),\dots,f_m(\boldsymbol{x})\big).
\end{equation}
Let $V:=\operatorname{span}_{\mathbbm{R}}\{F(\boldsymbol{x}):\boldsymbol{x}\in\mathbbm{R}^d\}\subseteq\mathbbm{R}^m$.
If $V\neq\mathbbm{R}^m$, then there exists $\boldsymbol{a}=(a_1,\dots,a_m)\neq \boldsymbol{0}$ with
$\langle \boldsymbol{a}, \boldsymbol{y} \rangle= \boldsymbol{0}$ for all $\boldsymbol{y}\in V$. In particular, $\boldsymbol{0}=\langle \boldsymbol{a}, \boldsymbol{F}(\boldsymbol{x}) \rangle =\sum_{i=1}^m \boldsymbol{a}_i f_i(\boldsymbol{x})$
for all $\boldsymbol{x}\in\mathbbm{R}^d$, contradicting the linear independence of $f_1,\dots,f_m$.
Hence $V=\mathbbm{R}^m$.

Therefore, we can choose points $\boldsymbol{x_1},\dots,\boldsymbol{x_m}\in\mathbbm{R}^d$ such that
$F(\boldsymbol{x_1}),\dots,F(\boldsymbol{x_m})$ are a basis of $\mathbbm{R}^m$ (equivalently, the
$m\times m$ matrix $\big(f_i(\boldsymbol{x_j})\big)_{i,j}$ is invertible).

Since $\mathbbm{R}^d$ is path connected, we can choose a continuous path $\boldsymbol{\gamma}:[0,T]\to\mathbbm{R}^d$ that visits these points in order:
there exist $0<t_1<\cdots<t_m<T$ with $\boldsymbol{\gamma}(t_j)=\boldsymbol{x_j}$ for each $j$.

Suppose towards a contradiction that $f_1\circ\boldsymbol{\gamma},\dots,f_m\circ\boldsymbol{\gamma}$ are
linearly dependent on $[0,T]$. Then there exists $\boldsymbol{a}=(a_1,\dots,a_m)\neq \boldsymbol{0}$ with
\begin{equation}
\sum_{i=1}^m a_i\, f_i(\boldsymbol{\gamma}(t)) = 0\qquad \forall \ t\in[0,T].
\end{equation}
Evaluating at $t=t_j$ gives $\sum_{i=1}^m a_i f_i(\boldsymbol{x_j})=0$ for $j=1,\dots,m$, i.e.
\begin{equation}
\begin{pmatrix}
f_1(\boldsymbol{x_1})&\cdots&f_m(\boldsymbol{x_1})\\
\vdots  &       &\vdots\\
f_1(\boldsymbol{x_m})&\cdots&f_m(\boldsymbol{x_m})
\end{pmatrix}
\begin{pmatrix}
a_1\\ \vdots\\ a_m
\end{pmatrix}
=0.
\end{equation}
Since the matrix is invertible, this forces $\boldsymbol{a}=\boldsymbol{0}$, a contradiction. Hence
$f_1\circ\gamma,\dots,f_m\circ\gamma$ are linearly independent on $[0,T]$. $\square$

Hence we can recast the cancelling results from the main text with a fixed path, yielding that we can find a path and a control function such that we do not cancel $f$ but $\{g_i\}_{i=1}^n$ if $f$ and $\{g_i\}_{i=1}^n$ are linearly independent. 
\subsection{Spatially and temporally correlated signals}\label{app:noisecanctime}
Here we show, that also for signals and noise signals with explicit time dependence, if the signal $f:\mathbbm{R}^d\times \mathbbm{R}\rightarrow \mathbbm{R}$ and the noise signals $\{ g_i : \mathbbm{R}^d \times \mathbbm{R} \to \mathbbm{R} \}_{i=1}^n$ are linearly independent there exists a path $\boldsymbol{\gamma}$ such that $f(\boldsymbol{\gamma}(\cdot),\cdot):[0,T]\rightarrow\mathbbm{R}$ and  $\{g_i(\boldsymbol{\gamma}(\cdot),\cdot):[0,T]\rightarrow\mathbbm{R}\}$ are linearly independent. 
\textbf{Proof:}
Define
\begin{equation}
h_0:=f,\qquad h_i:=g_i\quad (i=1,\dots,n).
\end{equation}
For each $(\boldsymbol{x},t)\in\mathbbm{R}^d\times\mathbbm{R}$ set
\begin{equation}
\boldsymbol{v}(\boldsymbol{x},t)
:=\big(h_0(\boldsymbol{x},t),h_1(\boldsymbol{x},t),\dots,h_{n}(\boldsymbol{x},t)\big)
\in\mathbbm{R}^{n+1}.
\end{equation}
The linear independence of $\{h_i\}_{i=0}^{n}$ means if for coefficients
$\boldsymbol{a}=(a_0,\dots,a_{n})$ satisfy
\begin{equation}
\langle \boldsymbol{a}, \boldsymbol{v}(\boldsymbol{x},t) \rangle=0,
\quad \forall \ (\boldsymbol{x},t) \in\mathbbm{R}^d\times\mathbbm{R},
\end{equation}
then $\boldsymbol{a}=0$. Equivalently, the linear span of
\begin{equation}
\big\{\boldsymbol{v}(\boldsymbol{x},t):(\boldsymbol{x},t)\in\mathbbm{R}^d\times\mathbbm{R}\big\}
\end{equation}
is all of $\mathbbm{R}^{n+1}$.
Hence we can choose $n+1$ points $(\boldsymbol{x}_k,t_k)\in\mathbbm{R}^d\times\mathbbm{R}$, $k=0,\dots,n$, with pairwise distinct $t_k$, such that the vectors
\begin{equation}
\boldsymbol{v}_k:=\boldsymbol{v}(\boldsymbol{x}_k,t_k)\in\mathbbm{R}^{n+1}
\end{equation}
are linearly independent. Let $V$ be the $(n+1)\times(n+1)$ matrix whose $k$-th row is $\boldsymbol{v}_k$. Then
\begin{equation}
\det V\neq 0.
\end{equation}
Define a continuous  path
\begin{equation}
\boldsymbol{\gamma}:[0,T]\to\mathbbm{R}^d
\end{equation}
such that, without loss of generality $0\le t_0<\cdots<t_n\le T$,
\begin{equation}
\boldsymbol{\gamma}(t_k)=\boldsymbol{x}_k\qquad (k=0,\dots,n).
\end{equation}
Suppose there exist constants $a_0,\dots,a_n$ with
\begin{equation}
\sum_{i=0}^{n} a_i\,h_i\big(\boldsymbol{\gamma}(t),t\big)= 0
\quad\text{for all }t\in[0,T].
\end{equation}
Evaluating at $t=t_k$ gives, for $k=0,\dots,n$,
\begin{equation}
\sum_{i=0}^{n} a_i\,h_i(\boldsymbol{x}_k,t_k)=0,
\end{equation}
i.e.
\begin{equation}
V\,\boldsymbol{a}=\boldsymbol{0},
\end{equation}
where $\boldsymbol{a}=(a_0,\dots,a_n)^\top$. Since $\det V\neq 0$, we have $\boldsymbol{a}=\boldsymbol{0}$. $\square$

\section{Cancelling of function basis families}\label{app:cancelling-noise-families}
Here, we consider that the signal and the noise are out of specific function families that constitute a basis for a certain function class, which can be motivated for example via looking at a series expansion of the signal function $f$, being interested in some term of the series, but treating the other terms (with unknown fluctuating coefficients) as noise parameters. For simplicity we consider here one dimensional functions $f:\mathbbm{R}^d\rightarrow\mathbbm{R}$ with $d=1$. 
\subsection{Trigonometric polynomials: Fourier series}
Consider a periodic function $f(x+P)=f(x)$, with period $P$, then the Fourier series of the function is given as
\begin{equation}
    f(x)=A_0+\sum_{n=1}^{\infty} A_i \cos(\frac{2 \pi n}{P}x-\phi_n).
\end{equation}
Let us consider the path $\gamma:[0,T]\rightarrow \mathbbm{R}$ with $\gamma(t)=\frac{tP}{T}$, obtaining the phase
\begin{equation}
    \Phi=\int_{0}^T f\left(\frac{tP}{T} \right) dt= \frac{T}{P}\int_{0}^P f(x) dx.
\end{equation}
We now introduce a control function $c$, implemented either purely through quantum control or through a modified velocity schedule combined with finite quantum control, yielding
\begin{equation}
    \Phi_c=\frac{T}{P} \int_0^P c(x)f(x)dx.
\end{equation}
Assuming we want to estimate coefficient $A_m$, choosing $c(x)=\sin\left( \frac{2 \pi m}{P}x-\phi_m \right)$ yields
\begin{equation}
    \Phi_c=\frac{T \pi}{P} A_m,
\end{equation}
as the trigonometric polynomials $ \{\cos(nx),\sin(mx) : n,m \in \mathbbm{N}_0 \}$ constitute an orthogonal basis in $L^2[- \pi, \pi]$.
\subsection{Polynomials: Taylor series}
Assume $f$ is analytical, so we can write it in terms of a Taylor series
\begin{equation}
    f=\sum_{n=0}^{\infty} a_n x^n.
\end{equation}
Now we want to estimate the $m$-th order of the series, while being insensitive to the lower orders. Note being insensitive to all orders, while retaining sensitivity to a specific order is impossible, as the polynomials are an over-complete basis of $L^2([0,1])$. We assume we move with constant velocity on a path $\gamma:[0,T] \rightarrow \mathbbm{R}$ with $\gamma(t)=vt$, yielding 
\begin{equation}
    \Phi_c=\int_0^T f(vt) dt = \frac{1}{v}\int_0^{vT} f(x) dx.
\end{equation}
To find the optimal control we need determine the optimal $L^1[0,vT]$-approximation of $x^m$ from the set $\mathrm{span}_{\mathbbm{R}}\{1,x,\dots,x^{m-1}\}$. The optimal polynomial is given by the $m$-th Chebyshev polynomial of second kind \cite{Tchebichef1874}
\begin{equation}
   U_m(x)=2\frac{\sin((m+1)\arccos(2 x-v T)/(v T))}{\sqrt{(v T-x)x}}\left(\frac{vT}{4}\right)^{m+1}
\end{equation}
minus the signal function 
\begin{equation}
    p^*=U_m(x)-x^m
\end{equation}
Hence the optimal control is given by $\mathrm{sign}(p^*)$ yielding the optimal residual sensitivity of \cite{Tchebichef1874}
\begin{equation}
    \Phi_c=\frac{4}{v}\left(\frac{vT}{4}\right)^{m+1}.
\end{equation}

In general if we want to be most sensitive to a certain order $l$ and insensitive to orders $I \subset \mathbbm{N}$ we have to find the best $L^1$-approximation $p^*$ and apply the techniques from the main text. 

\subsection{Orthogonal polynomials: Legendre polynomials}
The Legendre polynomials constitute an orthogonal basis of the piecewise continuos functions on the the interval $[-1,1]$. Hence given any piecewise continuos function on the interval $[-1,1]$, we obtain any expansion coefficient of the $m$-th Legendre polynomial 
\begin{equation}
P_m(x)
= \sum_{k=0}^{\left\lfloor m/2 \right\rfloor}
(-1)^k
\frac{(2m - 2k)!}{(m - k)!\,(m - 2k)!\,k!\,2^m}\,
x^{\,m - 2k}
\end{equation}
via choosing a path going through the whole domain and choosing as the control function the $m$-th Legendre polynomial. Formally, let $f:[-1,1] \rightarrow \mathbbm{R}$ be a piecewise continuos function, hence
\begin{equation}
    f(x)=\sum_{n=0}^{\infty} a_n P_n(x).
\end{equation}
Then,
\begin{equation}
    \Phi=\int_{0}^T f\left(\frac{-T+2t}{T}\right) dt= \frac{T}{2}\int_{-1}^1 f(x) dx 
\end{equation}
and 
\begin{equation}
    \Phi_c= \frac{T}{2} \int_{-1}^1 f(x)P_m(x)dx= a_m \frac{T}{2m+1}. 
\end{equation}

\section{Quantum Fisher Information for Time-Dependent Hamiltonians}\label{app:timedep}

We briefly derive the QFI bound and discuss its saturation via adaptive control following Ref.~\cite{Pang2017}.  
For a time-dependent Hamiltonian \(H_g(t)\), the probe evolves as 
\begin{equation}
\ket{\Psi_g(T)} = \underbrace{\exp(-i\int_0^T H_g(t)\,dt)}_{\equiv U_g(0\!\to\!T ) }\ket{\Psi_0}.
\end{equation}

The corresponding generator 
$h_g(T) = i[\partial_g U_g(0\!\to\!T)]U_g^\dag(0\!\to\!T)$
yields $I_g^{(Q)} = 4\,\mathrm{Var}[h_g(T)]_{\ket{\Psi_g(T)}}$.
The QFI is bounded by
\begin{equation}
    I_g^{(Q)} \le 
    \!\left[\!\int_0^T (\mu_{\max}(t)-\mu_{\min}(t))\,dt\!\right]^2,
\end{equation}
where $\mu_{\max,\min}(t)$ are the extremal eigenvalues of $\partial_g H_g(t)$.  
This bound is saturable via the optimal control Hamiltonian
\begin{equation}
\begin{aligned}
    H_c(t) = &\sum_k f_k(t)\ketbra{\Psi_k(t)} - H_g(t) 
    \\+ i&\sum_k \ket{\partial_t \Psi_k(t)}\bra{\Psi_k(t)},
\end{aligned}
\end{equation}
with $\ket{\Psi_k(t)}$ the eigenstates of the total Hamiltonian 
$H_{\mathrm{tot}}(t)=H_g(t)+H_c(t)$ and $f_k(t)$ real functions.  
In practice, $H_c(t)$ depends on the true parameter $g$ and is implemented using an estimate $g_c$, see \cite{Pang2017} for details.

\subsection{Example: Spatial Frequency Estimation}
For 
\begin{equation}
H_k(t) = -B[\cos(k\gamma(t))\sigma_x + 
\sin(k\gamma(t))\sigma_y],
\end{equation}
and $\gamma(t)=vt$,
$\partial_k H_k(t)=vtB[\sin(kvt)\sigma_x-\cos(kvt)\sigma_y]$
has extremal eigenvalues $\pm vtB$,
leading to
\begin{equation}
I_k^{(Q)} = \big[2B\!\int_0^T v t\,dt\big]^2 = B^2v^2T^4.
\end{equation}
For non-uniform motion \(v(t)\), the general expression becomes
\begin{equation}
I_k^{(Q)} = 4B^2\!\left[\!\int_0^T v(t)t\,dt\!\right]^2,
\end{equation}
and for uniform acceleration \(v(t)=v_0+at\),
\begin{equation}
I_k^{(Q)} = B^2T^4(v_0^2+\tfrac{4v_0a}{3}T+\tfrac{4a^2}{9}T^2),
\end{equation}
showing sextic scaling in $T$.
Expressing this in terms of total distance $L=v_0T+aT^2/2$ yields
\begin{equation}
I_k^{(Q)} = B^2T^2(L^2+\tfrac{1}{3}v_0T^3+\tfrac{7}{36}a^2T^4),
\end{equation}
which grows faster than for constant velocity.  
Maximizing the QFI corresponds to maximizing 
\(\int_0^T v(t)t\,dt = x_{\mathrm{sensor}}(T)T - \int_0^T x_{\mathrm{sensor}}(t)\,dt\), which is maximized if $|\int_0^T x_{\text{sensor}}(t)|$ is minimized. This expression is
minimized when motion is concentrated near the end of the interrogation, as the support of non-zero contributions is small.   
In the limit of instantaneous acceleration, 
\begin{equation}
I_k^{(Q)} = 4B^2T^2L^2.
\end{equation}

\subsection{General Moving-Sensor Hamiltonians}
For a signal Hamiltonian
\(H_{\mathrm{signal},\omega}(t)=\omega f(\boldsymbol{x}_{\mathrm{sensor}}(t))G\),
the spectral range of \(\partial_\omega H_{\mathrm{signal},\omega}(t)\)
gives
\[
I_\omega^{(Q)} = \|G\|_{\mathrm{spec}}^2
\!\left[\!\int_0^T f(\boldsymbol{\gamma}(t))\,dt\!\right]^2.
\]
For parameter-dependent signal functions 
\(H_{\mathrm{signal},k}(t)=f(k,\boldsymbol{x}_{\mathrm{sensor}}(t))G\),
the result generalizes to
\[
I_k^{(Q)} = \|G\|_{\mathrm{spec}}^2
\!\left[\!\int_0^T \partial_k f(k,\boldsymbol{\gamma}(t))\,dt\!\right]^2.
\]
For the general path dependent imprinting Hamiltonian
$H_{k}=H(k,\boldsymbol{\gamma}(t,k))$, using the chain rule we find 
\begin{equation}
\partial_k H_k 
=
\frac{\partial H}{\partial k}(k, \boldsymbol{\gamma})
+
\sum_{i=1}^{d}
\frac{\partial H}{\partial \boldsymbol{\gamma}_i}(k, \boldsymbol{\gamma})
\,
\frac{\partial \boldsymbol{\gamma}_i(k,t)}{\partial k},
\end{equation}
and taking the maximal and minimal eigenvalues we can obtain the optimal QFI.

\end{document}